\documentclass[aip,amsmath,amssymb,reprint]{revtex4-1}

\usepackage{graphicx}
\usepackage{dcolumn}
\usepackage{bm}
\usepackage[utf8]{inputenc}
\usepackage[T1]{fontenc}
\usepackage{mathptmx}
\usepackage[separate-uncertainty=true]{siunitx}

\begin{document}

\preprint{AIP/123-QED}

\title[]{Experimental analysis of the spin-orbit coupling dependence on the drift velocity of a spin packet}

\author{N. M . Kawahala}
    \affiliation{Instituto de Física, Universidade de São Paulo, São Paulo, SP 05508-090, Brazil}
    
\author{F. C. D. Moraes}
    \affiliation{Instituto de Física, Universidade de São Paulo, São Paulo, SP 05508-090, Brazil}
    
\author{G. M. Gusev}
    \affiliation{Instituto de Física, Universidade de São Paulo, São Paulo, SP 05508-090, Brazil}
    
\author{A. K. Bakarov}
    \affiliation{Institute of Semiconductor Physics and Novosibirsk State University, Novosibirsk 630090, Russia}
    
\author{F. G. G. Hernandez}
    \email{felixggh@if.usp.br}
    \affiliation{Instituto de Física, Universidade de São Paulo, São Paulo, SP 05508-090, Brazil}

\date{\today}

\begin{abstract}
Spin transport was studied in a two-dimensional electron gas hosted in a wide GaAs quantum well occupying two subbands. Using space and time Kerr rotation microscopy to image drifting spin packets under an in-plane accelerating electric field, optical injection and detection of spin polarization were achieved in a pump-probe configuration. The experimental data exhibited high spin mobility and long spin lifetimes allowing to obtain the spin-orbit fields as a function of the spin velocities. Surprisingly, above moderate electric fields of $\SI{0.4}{\volt/\centi\meter}$ with velocities higher than $\SI{2}{\micro\meter/\nano\second}$, we observed a dependence of both bulk and structure-related spin-orbit interactions on the velocity magnitude. A remarkable feature is the increase of the cubic Dresselhaus term to approximately half of the linear coupling when the velocity is raised to $\SI{10}{\micro\meter/\nano\second}$. In contrast, the Rashba coupling for both subbands decreases to about half of its value in the same range. These results yield new information for the application of drift models in spin-orbit fields and about limitations for the operation of spin transistors.
\end{abstract}

\maketitle

\section{Introduction}
Over the past few decades, the quest to build spintronic analogs to conventional charge-based electronic devices has motivated intense research \cite{wolf2001, zutic2004, awschalom2007, Wunderlich2010, egues2003}. Paramount to this search is the spin transistor, proposed by Datta and Das \cite{Datta1990}, that uses a gate-tunable Rashba spin-orbit interaction (SOI) \cite{Bychkov1984} for electric manipulation of the spin state inside a ballistic channel. Later studies included the Dresselhaus SOI \cite{Dresselhaus1955} so that a non-ballistic transistor robust against spin-independent scattering could be realized \cite{Schliemann2003, Ohno2008, Kunihashi2012, Kohda2017}. For instance, when both Rashba and Dresselhaus couplings (SOCs) have equal magnitudes ($\alpha=\beta$), a uniaxial spin-orbit field is formed and the spin polarization could be preserved during transport \cite{Bernevig2006, Koralek2009, Walser2012, Schliemann2017}.

In order to bring further robustness to more realistic spin transistors, we have to account for unwanted effects caused when applying in-plane electric fields inside the transistor channel as, for example, heating by the current. A recent study in a single subband system showed that heating of the electron system leads to a drift-induced enhancement of the cubic Dresselhaus SOI  \cite{Kunihashi2017}. It was also shown that carrier heating strongly increases the diffusion coefficient \cite{PhysRevB.99.125404}. Moreover, cubic fields introduced temporal oscillations of the spin polarization during transport by drift and cause spin dephasing \cite{PRL2016}. Alternately, new device architectures using external magnetic fields have been considered to overcome such dephasing in systems set to the persistent spin helix regime \cite{PhysRevB.101.155414}.

In this work, we are interested in the investigation of drift-induced SOCs modifications by exploring a two-dimensional electron gas (2DEG) hosted in a wide GaAs quantum well (QW) with two-occupied subbands. Previous studies in such multilayer systems shown high charge mobility and long spin  lifetimes \cite{Luengo-Kovac2017} as well as the possibility to generate current-induced spin polarization \cite{PRB2013, PRB2014, PRB2016}. Employing optical techniques for injection and detection of spin polarization, it was possible to image drifting spin packets and to obtain the spin mobility and spin-orbit field anisotropies. Further increasing the drift velocities, we have observed the enhancement of the cubic Dresselhaus SOI in agreement with the single subband case \cite{Kunihashi2017}. However, we have also observed an unexpected decrease in the Rashba SOI. These findings establish limitations to the assumption of constant spin-orbit couplings independent of the velocity range of the spin transistor operation.

\section{Experimental Measurements}
\begin{figure*}[ht]
    \centering
    \includegraphics[scale=0.5]{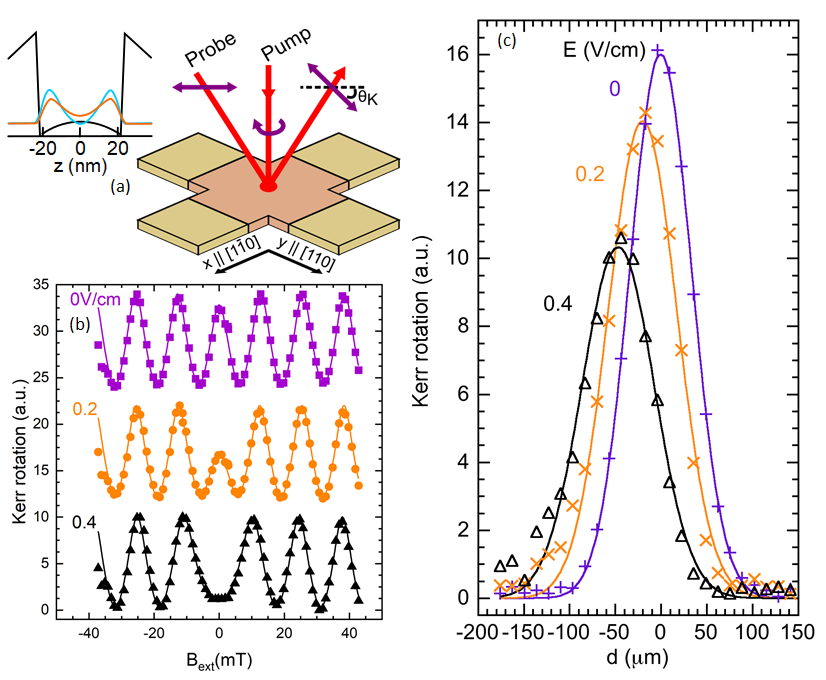}
    \caption{(a) Device geometry, contacts configuration, and a simple scheme of the pump-probe technique used for optical measurements. Also, the top left image shows the potential profile and the subbands charge density of the two-subband QW. (b) External magnetic field scan of the Kerr rotation signal measured at the spatial overlap of pump and probe beams for three different values of the in-plane electric field applied to the $y\text{-oriented}$ channel. Solid lines are fittings using Equation 1. (c) Spatial distribution of the spin polarization amplitudes, showing fittings with the expected Gaussian profiles (solid lines), and the displacement caused when applying the same electric fields displayed in (b).}
    \label{fig1}
\end{figure*}

The sample used in the investigations was a \SI{45}{\nano\meter} wide QW, symmetrically doped with Si, and grown in the $\left[001\right] (z)$ direction. The top left image in Figure \ref{fig1}(a) shows the calculated band profile for the QW. As the charge distribution experiences a soft barrier inside the well, caused by the Coulomb repulsion of the electrons, the electronic system is configured with symmetric and antisymmetric wave functions for the two lowest subbands with a separation of $\Delta_\text{SAS} = \SI{2}{\milli\electronvolt}$. The Shubnikov-de Hass oscillations provided the values of $n_1=\SI{3.7e11}{\centi\meter^{-1}}$ and $n_2=\SI{3.3e11}{\centi\meter^{-1}}$ for each subband density and the low-temperature charge mobility was given as $\mu_\text{c} = \SI{2.2e6}{\centi\meter\squared/\volt\second}$ \cite{Luengo-Kovac2017}. To induce the drift transport required for the spin-orbit field measurements, a cross-shaped device was fabricated with a width of $w=\SI{270}{\micro\meter}$ and channels along the $\left[1\bar{1}0\right] (x)$ and $\left[110\right] (y)$ directions, where lateral Ohmic contacts were deposited $l=\SI{500}{\micro\meter}$ apart so that in-plane voltages could be applied. A simple scheme of this device is also shown in Figure \ref{fig1}(a).

To perform spin polarization measurements using time-resolved Kerr rotation as a function of space and time, a mode-locked Ti:sapphire laser with a repetition rate of \SI{76}{\mega\hertz} tuned to \SI{819}{\nano\meter} was split into pump and probe beams. A photoelastic modulator was used to control the circular polarization of the pump beam and the optically generated spin polarization was then measured by analyzing the rotation of the reflected linearly polarized probe beam. The intensity of the probe beam was also modulated by an optical chopper for cascaded lock-in detection. The time delay of the probe pulse relative to the pump was controllable by mechanically adjusting the length of the pump path. The incidence of the probe beam was fixed to the center of the cross-shaped device while the pump beam could be moved with a scanning mirror, allowing to spatially map the drifting spin packets. All presented measurements were performed at $\SI{10}{\kelvin}$.

The spin-orbit fields are obtained by measuring the Kerr rotation signal when varying the strength of the external magnetic field ($B_\text{ext}$), applied in the plane of the QW, for a fixed space and time separation between pump and probe pulses. We model the dependence of the Kerr rotation angle $\Theta_\text{K}$ as following:

\begin{eqnarray}\label{eq:RSA}
    \Theta_\text{K}(B_\text{ext}) =
    \sum_{n} A_n\cos\Bigg[&&\frac{g\mu_\text{B}}{\hslash}\left(\Delta t + nt_\text{rep}\right)\times\nonumber\\
    &&\times\sqrt{\left(B_\text{ext}+B_{\text{SO},\parallel}\right)^2+B_{\text{SO},\perp}^2}\Bigg],
\end{eqnarray}

\noindent where $g$ is the electron g-factor, $\mu_\text{B}$ is the Bohr magneton, $\hslash$ is the reduced Planck constant, $\Delta t$ is the time separation between pump and probe pulses, $t_\text{rep}$ is the time interval between subsequent laser pulses and $B_{\text{SO},\parallel}$ ($B_{\text{SO},\perp}$) is  the component of the spin-orbit field parallel (perpendicular) to $B_\text{ext}$. Considering the device geometry, the resulting field is $B_{\text{SO},\perp}$ or $B_{\text{SO},\parallel}$ when the drifting electric-field is applied in the channel oriented parallel or perpendicular to $B_\text{ext}$, respectively. Next, we will refer to the magnitude of the spin-orbit field simply as $B_{\text{SO}}$ independently of its direction. For spin lifetimes that are longer than $t_\text{rep}$, the Kerr rotation measured at a given instant depends not only on the spin polarization injected by the last pump pulse but also on the remaining polarization from the previous ones. Although the contribution of each $n$-th pulse is accounted for by the sum in equation \eqref{eq:RSA}, only some of those terms are non-negligible since $A_n$ rapidly decreases as $n$ grows. All measurements presented in this work were made with a fixed long time delay of $\Delta t = \SI{12.9}{\nano\second}$ between injection and detection.

Figure \ref{fig1}(b) shows the Kerr rotation signal measured when scanning in a short range of $B_\text{ext}$ for three different values of the in-plane electric field ($E$) applied in the $y\text{-oriented}$ channel (set along $B_\text{ext}$). For each curve, the data was taken at the overlap position between pump and probe beams (peak amplitude in Figure \ref{fig1}(c)). Solid lines correspond to fittings of equation \eqref{eq:RSA} to the experimental data. Note that, while the components of $B_{\text{SO}}$ parallel and perpendicular to $B_\text{ext}$ shift laterally the data or decrease the magnitude of the center peak (as in Figure \ref{fig1}(b)), the g-factor changes the frequency of the peaks. Thus, the field  scan experiment determines the g factor and the spin-orbit field components separately as opposite to time-resolved measurements. 

Repeating field scans for several positions of pump-probe separation ($d$), the fitted amplitudes allow us to map the spin distribution and to track the drifting spin package. For instance, Figure \ref{fig1}(c) shows the extracted spin polarization amplitudes for the same electric field values in \ref{fig1}(b). The curves exhibit Gaussian profiles, expected as both pump and probe beams are also Gaussian, displaced from each other. Subtracting the center position for these distributions from the value at $E=0$ to calculate the spin package displacement $\Delta d$, the drift velocities can be computed as $\mathrm{v}=\Delta d/\Delta t$. 

\begin{figure}[ht]
    \centering
    \includegraphics[scale=0.52]{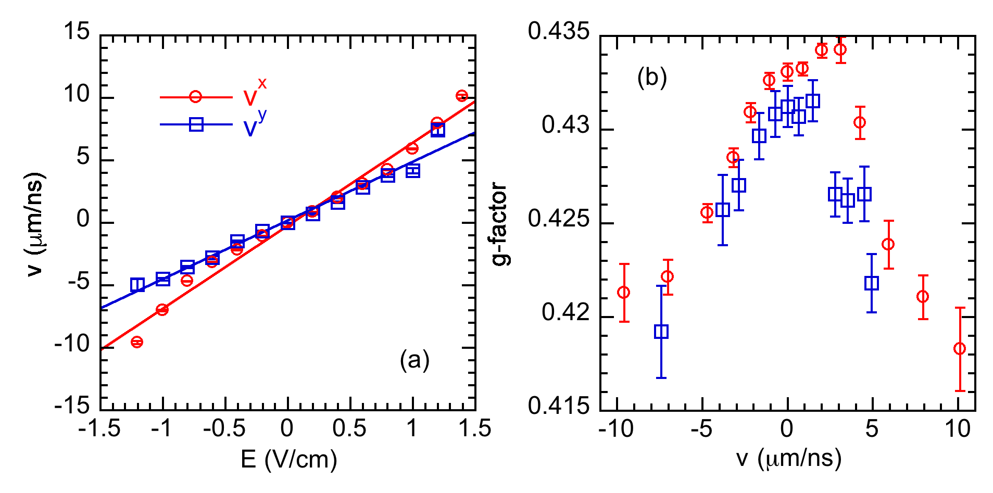}
    \caption{(a) Drift velocity $\mathrm{v}$ of the spin packet for in-plane electric fields applied in $x\text{-}$ and $y\text{-oriented}$ channels. The lines are linear fits from which spin mobility values were extracted. (b) Dependence of the g-factor with the drift velocities in each sample channel.}
    \label{fig2}
\end{figure}

Furthermore, the spin mobility ($\mu_\text{s}$) can also be obtained from the relation $\mathrm{v}(E)=\mu_\text{s}E$. Figure \ref{fig2}(a) shows the drift velocities obtained from the measurements performed for positive and negative values of $E$ applied in the $x$-(red circles) and $y$-oriented (blue squares) channels. The solid lines are linear fits with slope  $\mu_\text{s}^x=\SI{5.94(21)e5}{\centi\meter\squared/\volt\second}$ and $\mu_\text{s}^y=\SI{4.40(15)e5}{\centi\meter\squared/\volt\second}$. A similar mobility anisotropy was previously reported in this system and related to the existence of anisotropic spin-orbit fields. The spin mobility was found to be controlled by the SOCs setting the field along the direction perpendicular to the drift velocity \cite{Luengo-Kovac2017}.

Figure \ref{fig2}(b) shows the variation of the electron g-factor modulus with the in-plane electric field. Note that, for both channel directions, this modification is symmetric on the field polarity (velocity direction). A similar behavior was measured in InGaAs epilayers \cite{PhysRevB.91.201110}. However, the high mobility in our system lets us produce a stronger variation of 0.02 when increasing the drift velocity from zero to $\SI{10}{\micro\meter/\nano\second}$ in considerably smaller electric fields. The mechanism behind this g-factor dependence still requires investigation.

\begin{figure}[ht]
    \centering
    \includegraphics[scale=0.5]{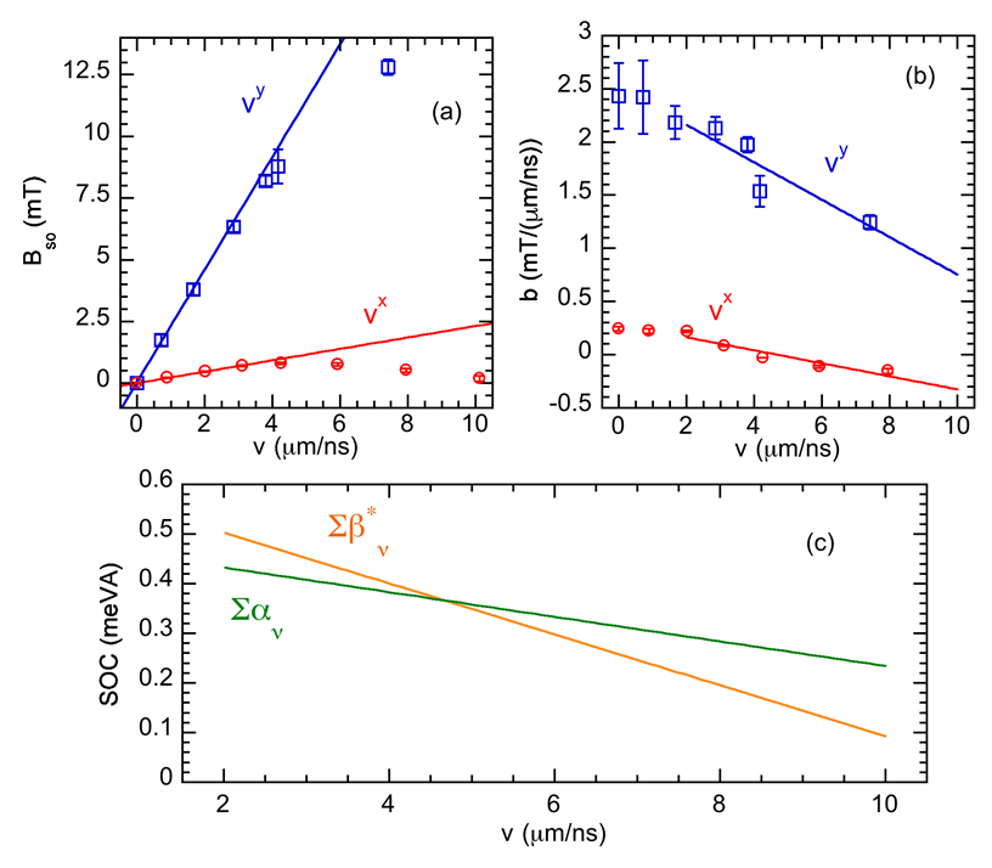}
    \caption{(a) Spin-orbit field $B_\text{SO}^{x(y)}$ as a function of the drift velocity $\mathrm{v}^{y(x)}$. (b) Ratio $B_\text{SO}/\mathrm{v}$, computed for each data point in (a), as a function of the drift velocities in each of the device channels. (c) Projection of the dependence of the SOCs on $\mathrm{v}$ computed by plugging the linear fits obtained in (b) into equations \eqref{eq:SOCs}.}
    \label{fig3}
\end{figure}

Next, we will focus on the dependence of the spin-orbit fields with an increasing drift velocity. As commented above, the $B_\text{ext}$ scans give an independent determination of the spin-orbit field components by fitting equation \eqref{eq:RSA}. Using the measured relation between $\mathrm{v}$ and $E$ in Figure \ref{fig2}(a), Figure \ref{fig3}(a) shows the $B_\text{SO}$ as a function of the drift velocities $\mathrm{v}^x$ (red circles) and $\mathrm{v}^y$ (blue squares) when $E$ is applied in the $x$- and $y$-oriented channels, respectively. These strong anisotropic fields can be expressed using a model \cite{Luengo-Kovac2017} in which the components of the spin-orbit field changes linearly with the transverse $\mathrm{v}$ as:

\begin{eqnarray}\label{model}
    &&\left<B_\text{SO}^x\right> = \left[\frac{m}{\hslash g\mu_\text{B}}\sum_{\nu=1}^2\Big(+\alpha_\nu+\beta_\nu^*\Big)\right]\mathrm{v}^y,\nonumber\\
    &&\left<B_\text{SO}^y\right> = \left[\frac{m}{\hslash g\mu_\text{B}}\sum_{\nu=1}^2\Big(-\alpha_\nu+\beta_\nu^*\Big)\right]\mathrm{v}^x,
\end{eqnarray}

\noindent where $\nu$ is the subband index $={1,2}$, $m=\SI{0.067}m_0$ is the effective electron mass for GaAs, $\alpha_\nu$ is the Rashba SOC for each subband and $\beta_\nu^*= \beta_{1,\nu}-2\beta_{3,\nu}$ depends on the linear ($\beta_{1,\nu}$) and cubic Dresselhaus ($\beta_{3,\nu}$) SOCs. The SOCs sum over the subband index indicates that the dynamics is governed by the average spin-orbit fields because the studied sample has an electron system with a strong intersubband scattering \cite{PhysRevB.95.125119}.

Figure \ref{fig3}(a) displays a non-monotonic function for $B_\text{SO}$ which is opposite to the prediction of equation \eqref{model}. The solid line is a linear fit to the data points in the low-velocity range. The departure from the model means that the SOCs inside the brackets change with an increasing drift velocity. Those parameters are graphically represented by the local curve slope and can be inferred by the ratios $b^y=B_\text{SO}^x/\mathrm{v}^y$ and $b^x=B_\text{SO}^y/\mathrm{v}^x$, plotted in Figure \ref{fig3}(b) as a function of $\mathrm{v}$. Clearly, $b^x$ and $b^y$ (and the SOCs) remain constant only up to approximately $\SI{2}{\micro\meter/\nano\second}$ followed by a decrease with increasing $\mathrm{v}$, that is stronger for $b^y$ than for $b^x$.

Finally, to translate the spin-orbit field dependence into the variation of the SOCs, we assumed a linear relation between $b$ and $\mathrm{v}$ in the high-speed regime ($\mathrm{v}>\SI{2}{\micro\meter/\nano\second}$). Using that relation, illustrated by the fit lines in Figure \ref{fig3}(b), we can isolate the Rashba and Dresselhaus terms from $b$ using the following equations:

\begin{equation}\label{eq:SOCs}
    \sum_{\nu=1}^2\beta_\nu^* = \frac{\hslash g\mu_\text{B}}{2m}\Big(b^y+b^x\Big), \quad \sum_{\nu=1}^2\alpha_\nu = \frac{\hslash g\mu_\text{B}}{2m}\Big(b^y-b^x\Big),
\end{equation}

\noindent constructed from the definition of the ratios $b^x$ and $b^y$ above.

Projections from equation \eqref{eq:SOCs} are shown in Figure \ref{fig3}(c). The sum of the Rashba coefficients decreases to half of its value at low velocities ($\SI{0.2}{\milli\electronvolt\angstrom}$) when increasing $\mathrm{v}$ up to $\SI{10}{\micro\meter/\nano\second}$. As $\alpha$ depends on the QW symmetry, the observed dependence with an in-plane electric field is unexpected and may be related to the same mechanism affecting the g-factor, which is still not understood. On the other hand, the variation of $\beta^*=\beta_{1}-2\beta_{3}$ is even stronger. Considering that the linear Dresselhaus coefficient depends on the QW width, it is expected that the measured modification of $\beta^*$ reflects exclusively an increase of $\beta_3$ as $\mathrm{v}$ grows. Similar enhancement of the cubic Dresselhaus SOC was previously discussed in single subband systems and associated with heating due to the high currents used to induce high drift velocities, as $\beta_3$ depends on the average kinetic energy \cite{Kunihashi2017,PhysRevB.99.125404}. Remarkably, we measured a strong enhancement of the cubic Dresselhaus term up to approximately half of the linear coupling when  $\beta^*\sim0$ at $\SI{10}{\micro\meter/\nano\second}$.

\section{Conclusions}
Cubic spin-orbit fields impose relevant constrains in spin transistor proposals that target extended coherence, thus demanding particular attention. Here, we addressed this issue in the investigation of a 2DEG confined in a GaAs QW with two occupied subbands. Applying in-plane electric fields, that enabled drift transport of spin packets along the device channels, we measured velocities as high as $\SI{10}{\micro\meter/\nano\second}$ and spin mobility in the range of $\SI{e5}{\centi\meter\squared/\volt\second}$. We observed the dependence of the spin-orbit couplings on the spin packet drift velocity. We found two regimes: i) For low velocities ($\mathrm{v}<\SI{2}{\micro\meter/\nano\second}$), the SOCs are independent of the drift velocity and the spin-orbit fields increase with increasing velocity, ii) For high velocities ($\mathrm{v}>\SI{2}{\micro\meter/\nano\second}$), the Rashba SOC decreases while the cubic Dresselhaus SOC increases and the spin-orbit fields become weaker with increasing velocity. Our findings indicate that limitations on the transport velocities should be considered when implementing spin transistors in multilayer systems.

\begin{acknowledgments}
We acknowledge financial support from São Paulo Research Foundation (FAPESP) Grants No. 2009/15007-5, No. 2013/03450-7, No. 2014/25981-7, No. 2015/16191-5, No. 2016/50018-1, and No. 2018/06142-5, Grants No. 301258/2017-1 and No. 131114/2017-4 of the National Council for Scientific and Technological Development (CNPq).
\end{acknowledgments}

\nocite{*}
\bibliography{aipsamp}
\end{document}